\documentclass[12pt,aps,eqsecnum,nofootinbib,floatfix]{revtex4}
\usepackage{bbding}
\usepackage{pifont}
\usepackage{subfigure}
\usepackage{epsfig}
\usepackage{array}%\usepackage{CJK}
\usepackage{amsmath}
\usepackage{amsfonts}
\usepackage{amssymb}
\usepackage{color}
\usepackage{graphicx}
\usepackage{mathrsfs}
\usepackage{multirow}

\def\be{\begin{equation}}
\def\ee{\end{equation}}
\def\ba{\begin{eqnarray}}
\def\ea{\end{eqnarray}}
\def\no{\nonumber}

\newcommand{\omits}[1]{}

\begin{document}

\title{Maxwell's equal area law for black holes \\
in power Maxwell
invariant}
\author{Huai-Fan Li$^{1,2}$\footnote{email address: huaifan.li@stu.xjtu.edu.cn, Huai-Fan Li}, Xiong-ying Guo$^{1,2}$, Hui-Hua Zhao$^{1,2}$ and Ren Zhao$^{1,2}$}
\affiliation{$^1$Institute of Theoretical Physics, Shanxi Datong University,
Datong 037009, China \\
$^2$Department of Physics, Shanxi Datong University,
Datong 037009, China}

\begin{abstract}
In this paper, we consider the phase transition of black hole in power Maxwell
invariant by means of Maxwell's equal area law. First, we review and study the analogy
of nonlinear charged black hole solutions with the Van der Waals gas-liquid system in
the extended phase space, and obtain isothermal $P$-$v$ diagram. Then, using the Maxwell's equal area law
we study the phase transition of AdS black hole with different temperatures.
Finally, we extend the method to the black hole in the canonical (grand canonical)
ensemble in which charge (potential) is fixed at
infinity. Interestingly, we find the phase transition occurs in
the both ensembles. We also study the effect of the
parameters of the black hole on the two-phase coexistence.
The results show that the black hole may go through a small-large
phase transition similar to those of usual non-gravity thermodynamic systems.
\end{abstract}

\maketitle

%%%%%%%%%%%%%%%%%%%%%%%%%%%%%%%%%%%%%

\section{Introduction}

In recent years, the cosmological constant in $n$-dimensional AdS and dS
spacetime has been regarded as pressure of black hole thermodynamic system.
The $(P,v)$ critical behaviors in AdS and dS black holes have
been extensively studied~\cite{David,Dola,Guna,Frass,David1,Altami,Altami1,Altami2,Zhao,Majhi,Majhi1,Majhi2,
Zhao1,Zhao2,Ma,Zhang,Ma1,Ma2,Heidi,Heidi1,Heidi2,
Heidi3,Heidi4,Heidi5,Heidi6,Heidi7,Arci,Azreg,Azreg1,Mo,Mo1,Mo2,Mo3,Mo4,Mo5,Lala,Wei,
Suresh,Mans,Thar,Niu,Ma3,Cai,Zou,Zou1,Li,Wei1,Poshteh,Xu,Liu,Xu1}.
It shows that black holes also have the standard thermodynamic quantities,
such as temperature, entropy, even possess abundant phase structures like
the Hawking-Page phase transition and the critical phenomena similar to ones in
the ordinary thermodynamic system. What is more interesting is the research on
charged, non-rotating RN-AdS black hole, which shows that there exists a phase transition
similar to the van der Waals-Maxwell gas-liquid phase transition~\cite{David,Dola,Guna,Frass,David1,Altami,Altami1,Altami2,
Zhao,Majhi,Majhi1,Majhi2,Zhao1,Zhao2,Ma,Zhang,Ma1,Ma2,Heidi,Heidi1,Heidi2,
Heidi3,Heidi4,Heidi5,Heidi6,Heidi7,Arci,Azreg,Azreg1,Mo,Mo1,Mo2,Mo3,Mo4,Mo5,Lala,Wei,
Suresh,Mans,Thar,Niu,Ma3,Cai,Zou,Zou1,Li,Wei1,Poshteh,Xu,Liu,Xu1}.

The isotherms in $(P,v)$ diagrams of AdS black hole in
Ref.~\cite{David,Dola,Guna,Frass,David1,Altami,Altami1,Altami2,Zhao,Majhi,Majhi1,Majhi2}
show there exists thermodynamic unstable region with
$\partial P/\partial v>0$ when temperature is
below critical temperature and the negative pressure emerges when
temperature is below a certain value. This situation also exists in
van der Waals-Maxwell gas-liquid system, which has been resolved by
Maxwell's equal area law~\cite{Spal,Zhao3,Spal2}. At this point,
it is worth mentioning that the Maxwell equal area construction can be
equally applied in the $(P,v)$ plane at constant temperature. This has
been done in ~\cite{Spal,Spal2} with interesting
results: i) the equal area law can be analytically solved; ii)
the unphysical negative specific heat region is cut off; iii)
a new black hole phase structure emerges; iv) the role of the Van der Waals
un-shrinkable molecular volume taken by the extremal black hole
configuration, thus justifying its stability. So, we hope that the Maxwell's equal
area laws can help us to find more phenomenon in the thermodynamics of black hole.

By this observations one may find it is worthwhile to study the
effects of nonlinear electrodynamics (NLEDs) on phase transition
of black holes in the extended phase space. In this direction, the
effects of nonlinear electromagnetic field of static and rotating
AdS black holes in the extended phase space have been analyzed~\cite{Heidi}
. In the last five years, a class of NLEDs has been
introduced, the so-called power Maxwell invariant (PMI) field
. The PMI field is significantly
richer than that of the Maxwell field, and in the special case
($s=1$) it reduces to linear electromagnetic source. The black
hole solutions of the Einstein-PMI theory and their interesting
thermodynamics and geometric properties have been examined before
.

In this paper, using the Maxwell's equal area law,
we establish a phase transition process in charged AdS black holes with PMI,
where the issues about unstable states and negative pressure
are resolved. By studying the phase transition process, we acquire the two-phase
equilibrium properties including the $P-v$ phase diagram.Using the Maxwell's equal area law
we study the phase transition of AdS black hole with different temperature.
Finally, we extend the method to the black hole in the (grand canonical)
canonical ensemble in which (potential) charge is fixed at
infinity. Interestingly, we find the phase transition occurs in
the both of canonical and grand canonical ensembles. We also study the effect of the
parameters of the black hole on the two phases coexistence.
The results show the phase transition below critical temperature is of the first order but phase transition
at critical point belongs to the continuous one.

The paper is organized as follows: In Sec. \ref{BH}, we review and consider
spherically symmetric black hole solutions of Einstein gravity in
the presence of the PMI source. Regarding the cosmological
constant as thermodynamic pressure, we study thermodynamic
properties and obtain Smarr's mass relation. In Sec. \ref{TP},
by Maxwell's equal area law the phase transition processes at certain
temperatures are obtained and the boundary of two phase equilibrium region are depicted in $P-v$
diagram for a charged AdS black hole with PMI. Then
some parameters of the black hole are analyzed to find the relevance
with the two-phase equilibrium. In Sec. \ref{Cano}, we consider the possibility of the
phase transition in the BTZ-like black hole and the grand canonical ensemble and find that in
contrast to RN black holes, the phase transition occurs. Finally, we finish this work with some concluding
remarks.

\section{Extended phase-space thermodynamics of black holes with PMI source}

\label{BH}

The bulk action of Einstein-PMI gravity has the following form
\begin{equation}
\label{Action}
I_{b}=-\frac{1}{16\pi }\int_{M}d^{n+1}x\sqrt{-g}\left( R+\frac{n(n-1)}{l^{2}}%
+\mathcal{L}_{PMI}\right) ,
\end{equation}
where $\mathcal{F}=F_{\mu \nu}F^{\mu \nu }$. Expanding the PMI Lagrangian near the
linear Maxwell case $(s \to 1)$, one can obtain
\begin{equation}
\label{eq2}
\mathcal{L}_{PMI} = ( - \mathcal{F})^s \to \mathcal{L}_{Max} + o(s - 1),
\end{equation}
where $\mathcal{L}_{Max} = - \mathcal{F}$ is the Maxwell Lagrangian.We consider a spherically symmetric spacetime as
\begin{equation}
ds^{2}=-f(r)dt^{2}+\frac{dr^{2}}{f(r)}+r^{2}d\Omega _{d-2}^{2},
\label{Metric}
\end{equation}
where $d\Omega _{d-2}^{2}$ stands for the standard element on $S^{d-2}$.
Considering the field equations following from the variation of the bulk
action with Eq. (\ref{Metric}), one can show that the metric function $f(r)$%
, gauge potential one--form $A$ and electromagnetic field two--form $F$ are
given by~\cite{Heidi,Hass,Hass2,Hendi1,Hendi2,Hendi3,Hendi4}
\begin{eqnarray}
f(r) &=&1+\frac{r^{2}}{l^{2}}-\frac{m}{r^{n-2}}+\frac{(2s-1)^{2}\left( \frac{%
(n-1)(2s-n)^{2}q^{2}}{(n-2)(2s-1)^{2}}\right) ^{s}}{%
(n-1)(n-2s)r^{2(ns-3s+1)/(2s-1)}},  \label{metfunction} \\
A &=&-\sqrt{\frac{n-1}{2(n-2)}}qr^{(2s-n)/(2s-1)}dt,  \label{A} \\
F&=&dA.  \label{dA}
\end{eqnarray}
The power $s \neq n/2$ denotes the nonlinearity parameter of the source
which is restricted to $s>1/2$. In the above expression, $m$ appears
as an integration constant and is related to the Arnowitt-Deser-Misnsr(ADM) mass of
the black hole. According to the definition of mass due to Abbott and Deser, the mass
of the soulution (\ref{metfunction}) is~\cite{Heidi}
\begin{eqnarray}
M &=&\frac{\omega _{n-1}}{16\pi }(n-1)m,  \label{Mass}
\end{eqnarray}
the electric charge is
\begin{eqnarray}
Q &=&\frac{\sqrt{2} (2s-1)s\; \omega _{n-1}}{8\pi }\left( \frac{n-1}{n-2}%
\right) ^{s-1/2}\left( \frac{\left( n-2s\right) q}{2s-1}\right) ^{2s-1},
\label{Charge}
\end{eqnarray}
where $\omega _{n-1}$ represents the volume of constant curvature
hypersurface described by $d\Omega _{n - 1}^2$,
\begin{equation}
\omega _{n-1}=\frac{2\pi ^{\frac{n}{2}}}{\Gamma \left( \frac{n}{2}\right) }.
\label{Omega}
\end{equation}
The cosmological constant is related to spacetime dimension $n$ by
\begin{equation}
\label{eq1}
\Lambda = - \frac{n(n - 1)}{2l^2},
\end{equation}
The Hawking temperature of the black hole on the outer horizon $r_ + $ can
be calculated using the relation
\begin{equation}
\label{eq2}
T = \frac{\kappa }{2\pi } = \frac{f'(r_ + )}{4\pi },
\end{equation}
where $\kappa $ is the surface gravity. The, one can easily show that
\begin{equation}
\label{eq3}
T = \frac{n - 2}{4\pi r_ + }\left( {1 + \frac{n}{n - 2}\frac{r_ + ^2 }{l^2}
- \frac{(2s - 1)}{(n - 1)(n - 2)r_ + ^{2(ns - 3s + 1) / (2s - 1)} }\left(
{\frac{(n - 1)(2s - n)^2q^2}{(n - 2)(2s - 1)^2}} \right)^s} \right),
\end{equation}
with $r_ + $ denotes the radius of the event horizon which is the largest
root of $f(r_ + ) = 0$. The electric potential $\Phi $, measured at infinity
with respect to the horizon while the black hole entropy $S$, determined from
the area law. It is easy to show that
\begin{equation}
\label{eq4}
\Phi = \sqrt {\frac{n - 1}{2(n - 2)}} \frac{q}{r_ + ^{(n - 2s) / (2s - 1)}
},
\end{equation}
\begin{equation}
\label{eq5}
S = \frac{\omega _{n - 1} r_ + ^{n - 1} }{4}.
\end{equation}
One may then regard the parameters $S$, $Q$, and $P$ as a complete set of
extensive parameters for the mass $M(S,Q,P)$ and define the intensive
parameters conjugate to $S$, $Q$, and $P$. These quantities are the
temperature, the electric potential and volume.
\begin{equation}
\label{eq6}
T = \left( {\frac{\partial M}{\partial S}} \right)_{Q,P} ,
\quad
U = \left( {\frac{\partial M}{\partial Q}} \right)_{S,P} ,
\quad
V = \left( {\frac{\partial M}{\partial P}} \right)_{Q,S} ,
\end{equation}
where~\cite{Cvetic}
\begin{equation}
\label{eq7}
P = \frac{n(n - 1)}{16\pi l^2},
\quad
V = \frac{\omega _{n - 1} }{n}r_ + ^n .
\end{equation}
It is a matter of straightforward calculation to show that the quantities
calculated by Eq. (\ref{eq7}) for the temperature, and the electric potential
coincide with Eqs. (\ref{eq3}) and (\ref{eq5}). Thus, the thermodynamics quantities
satisfy the first law of thermodynamics
\begin{equation}
\label{eq8}
dM = TdS + UdQ + VdP.
\end{equation}
From the above calculation, the thermodynamic quantities energy $M$, entropy $S$,
temperature $T$, volume $V$, pressure $P$, electric potential $U$ and electric charged $Q$ satisfy the Smarr
formula:
\begin{equation}
\label{eq9}
M = \frac{(n - 1)}{(n - 2)}TS + \frac{ns - 3s + 1}{s(2s - 1)(n - 2)}\Phi Q -
\frac{2}{n - 2}VP.
\end{equation}
In what follows we concentrate on analyzing the phase transition of the
black hole with PMI source system in the extended phase space while we treat
the black hole charge $Q$ as a fixed external parameter,
or the cosmical constant is a invariable parameters, not a thermodynamic variable. We shall find
that an even more remarkable coincidence with the Van der Waals fluid is
realized in this case.

Using the Eqs. (\ref{eq3}) and (\ref{eq7}) for a fixed charge Q, one may obtain the
equation of state, $P(v,T)$
\begin{equation}
\label{eq10}
P = \frac{T}{v}
 - \frac{(n - 2)}{\pi (n - 1)v^2}
 + \frac{1}{16\pi }\frac{kq^{2s}}{v^{2s(n - 1) / (2s - 1)}},
\end{equation}
\begin{equation}
\label{eq11}
k = \frac{4^{2s(n - 1) / (2s - 1)}(2s - 1)\left( {\textstyle{{(n - 1)(2s -
n)^2} \over {(n - 2)(2s - 1)^2}}} \right)^s}{(n - 1)^{2s(n - 1) / (2s -
1)}},
\end{equation}
where
\begin{equation}
\label{eq20}
v = \frac{4}{(n - 1)}r_ + ,
\end{equation}
is specific volume.

\begin{figure}[tbp]
$%
\begin{array}{cc}
\epsfxsize=5cm \epsffile{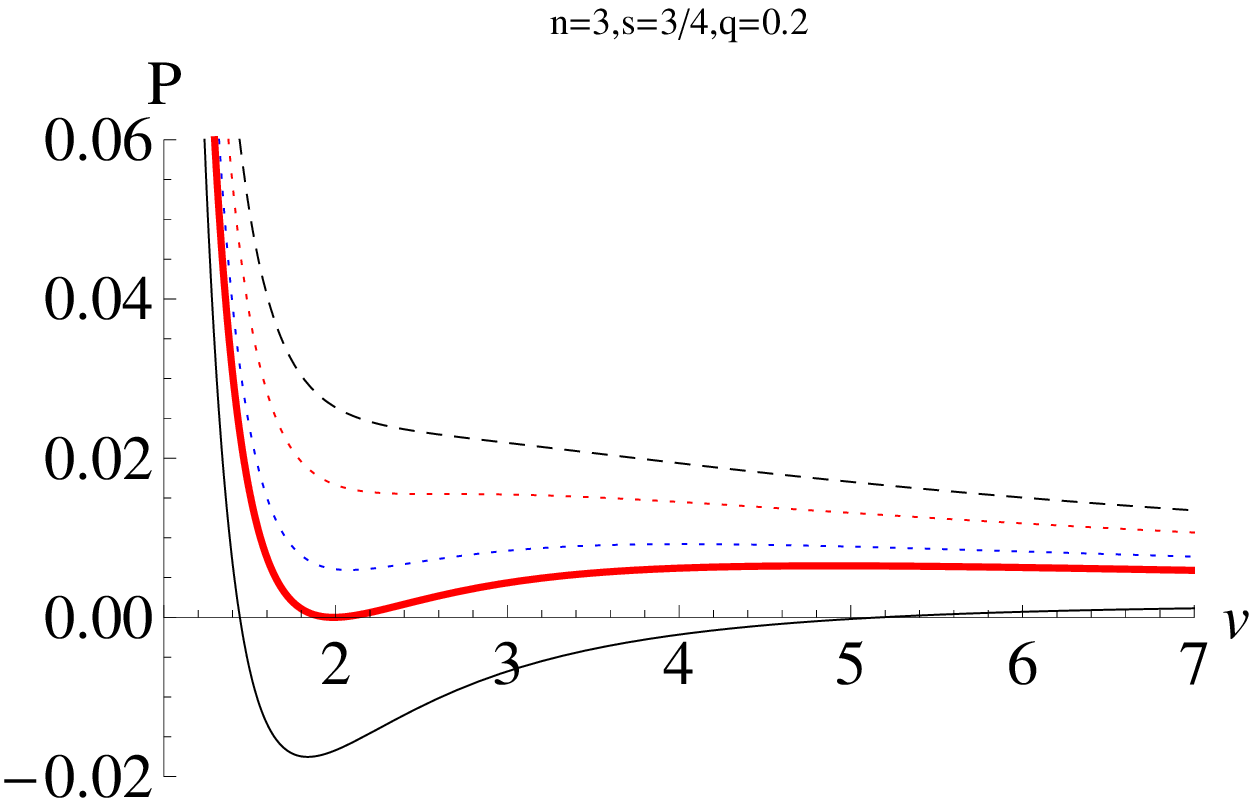} & \epsfxsize=5cm %
\epsffile{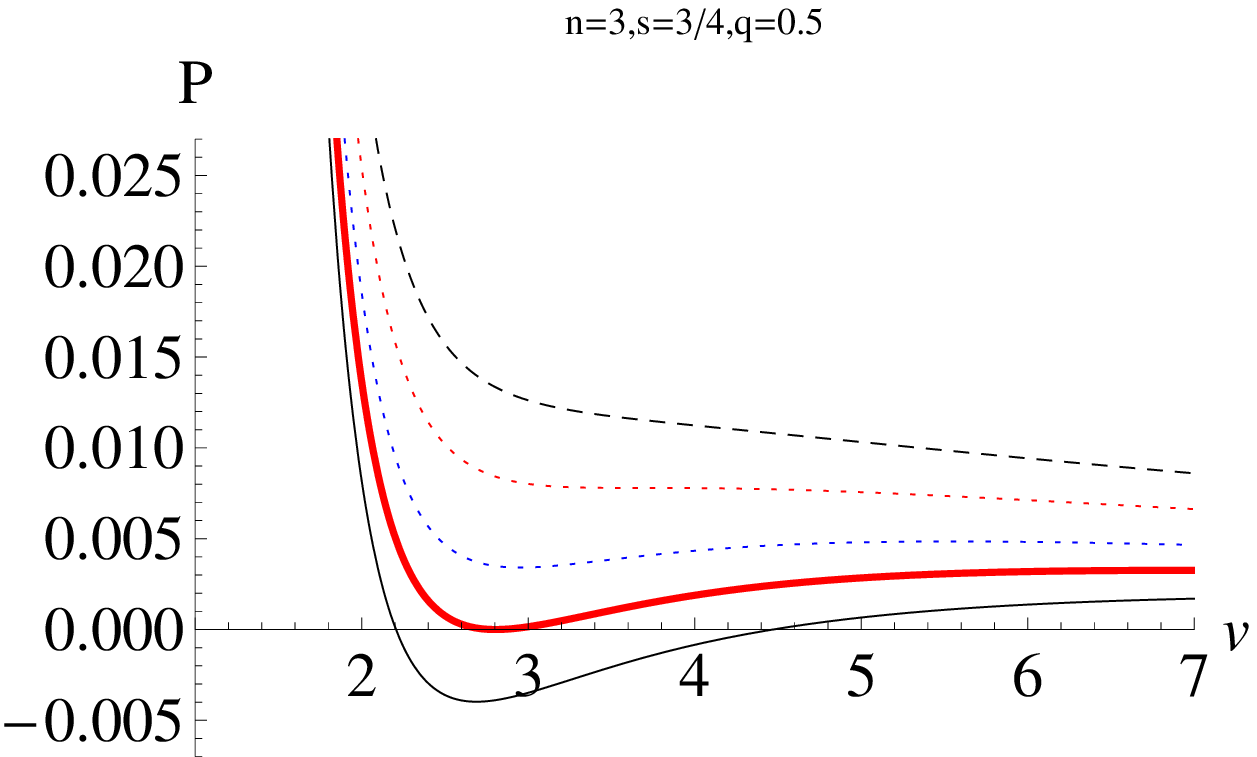}\epsfxsize=5cm %
\epsffile{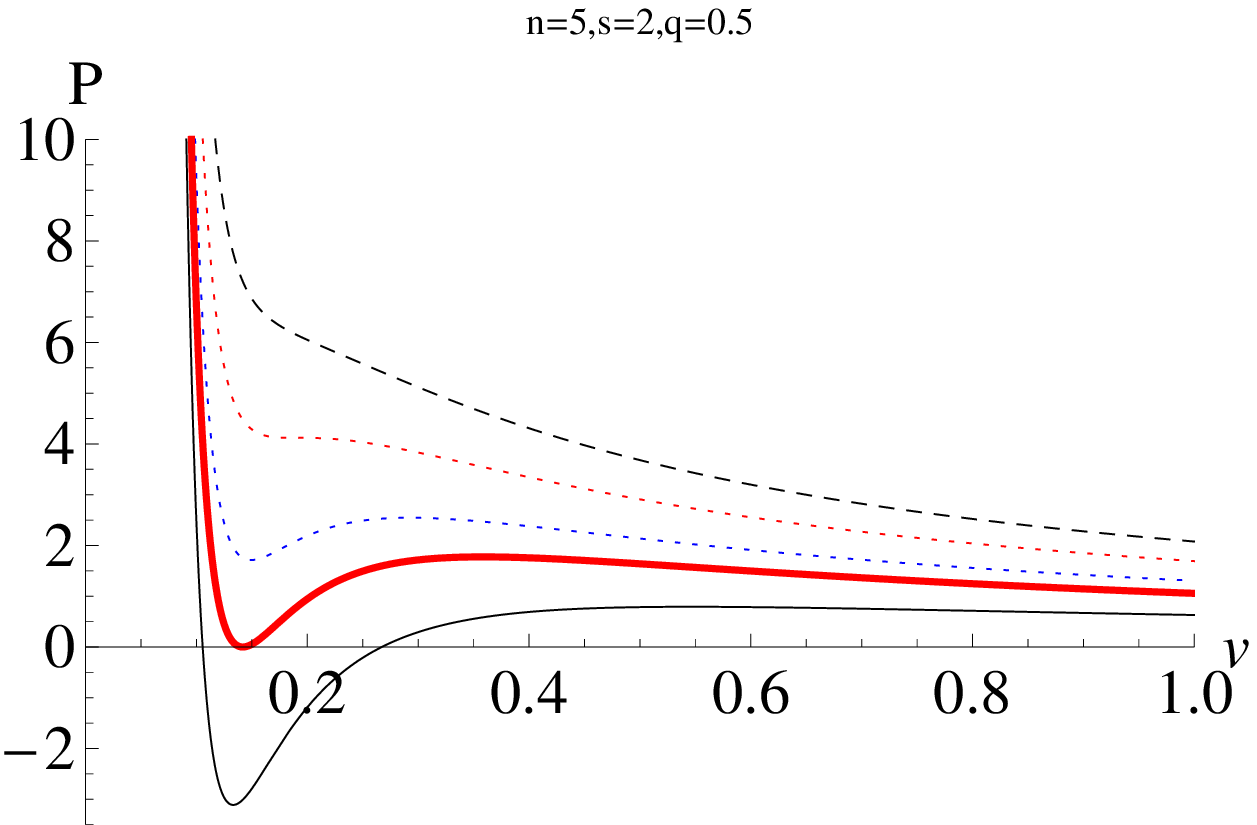}%
\end{array}
$%
\caption{Isotherms in $P-v$ diagrams of charged AdS black holes in PMI
in AdS spacetime. The temperature of isotherms decreases from top to bottom.}
\label{Pvn}
\end{figure}

In Fig.\ref{Pvn} we plot the isotherms in $P-v$ diagrams at different
dimension $n$, nonlinear parameters $s$, charge $q$. One can see from
Fig.\ref{Pvn} that there are thermodynamic unstable segments with $\partial P/\partial v>0$ on
the isotherms when temperature $T<T_c $, where $T_c $ is critical
temperature. When the temperature $T=T_0$, there is a point of
intersection between the isotherms and the horizontal $v$ axis.
 And the negative pressure emerges when temperature is
below certain value $T_0$. Using the above equation, $T_0$ and the corresponding
specific volume $v_0$ can be derived,
\begin{equation}
\label{eq12}
T_0 = \frac{(n - 2)}{\pi (n - 1)v_0 } + \frac{kq^{2s}}{16\pi
v_0^{\textstyle{{2s(n - 1)} \over {2s - 1}} - 1} },
\quad
v_0 =(\frac{kq^{2s}(2sn - 1)(n
- 1)}{16(2s - 1)(n - 2)})^\frac{1}{\textstyle{{2s(n - 1)} \over {2s - 1}} - 2}.
\end{equation}

\section{two-Phase equilibrium and Maxwell equal area law}
\label{TP}
The state equation of the charged black hole with PMI is exhibited by the
isotherms in Fig.\ref{Pvn}, in which the thermodynamic unstable states
with $\partial P/\partial v>0$ will lead to the system
expansion or contraction automatically and the negative pressure
situation have no physical meaning. The cases occur also in van der
Waals equation but they have been resolved by Maxwell equal area law.

We extend the Maxwell equal area law to $n$-dimensional charged
AdS black hole with PMI to establish an phase transition process of the black
hole thermodynamic system. On the isotherm with temperature $T_0 $ in $P-v$
diagram, the two points $\left( {P_0 ,\;v_1 } \right)$ and $\left( {P_0
,\;v_2 } \right)$ meet the Maxwell equal area law,
\begin{equation}
\label{eq13}
P_0 (v_2 -v_1 )=\int\limits_{v_1 }^{v_2 } {Pdv} ,
\end{equation}
which results in
\begin{equation}
\label{eq14}
P_0 (v_2 - v_1 ) = T_0 \ln \left( {\frac{v_2 }{v_1 }} \right) - A\left(
{\frac{1}{v_1 } - \frac{1}{v_2 }} \right)
 + \frac{B}{d - 1}\left( {\frac{1}{v_1^{d - 1} } - \frac{1}{v_2^{d - 1}
}} \right),
\end{equation}
where the two points $\left( {P_0 ,\;v_1 } \right)$ and $\left( {P_0 ,\;v_2
} \right)$ are seen as endpoints of isothermal phase transition. Considering
\begin{equation}
\label{eq15}
P_0 = \frac{T_0 }{v_1 } - \frac{A}{v_1^2 } + \frac{B}{v_1^d },
\quad
P_0 = \frac{T_0 }{v_2 } - \frac{A}{v_2^2 } + \frac{B}{v_2^d },
\end{equation}
from the eq.(\ref{eq15}), we can get
\begin{equation}
\label{eq16}
0 = T_0 \left( {\frac{1}{v_1 } - \frac{1}{v_2 }} \right) - A\left(
{\frac{1}{v_1^2 } - \frac{1}{v_2^2 }} \right) + B\left( {\frac{1}{v_1^d }
- \frac{1}{v_2^d }} \right),
\end{equation}
\begin{equation}
\label{eq17}
2P_0 = T_0 \left( {\frac{1}{v_1 } + \frac{1}{v_2 }} \right) - A\left(
{\frac{1}{v_1^2 } + \frac{1}{v_2^2 }} \right) + B\left( {\frac{1}{v_1^d }
+ \frac{1}{v_2^d }} \right) ,
\end{equation}
where
$$
A = \frac{(n - 2)}{\pi (n - 1)},
\quad
B = \frac{kq^{2s}}{16\pi },
\quad
d = \frac{2s(n - 1)}{2s - 1}.
$$
From the eqs.(\ref{eq15}), (\ref{eq16}) and (\ref{eq17}), we can obtain
\begin{equation}
\label{eq18}
T_0 v_2^{d - 1} x^{d - 1} = Av_2^{d - 2} x^{d - 2}(1 + x) - B\frac{1 -
x^d}{1 - x},
\end{equation}
and
\begin{equation}
\label{eq19}
v_2^{d - 2} = \frac{B}{A}\frac{d(1 - x^{d - 1})(1 - x) + (d - 1)(1 - x^d)\ln
x}{x^{d - 2}(d - 1)(1 - x)\left( {2(1 - x) + (1 + x)\ln x} \right)} = f(x),
\end{equation}
Substituting (\ref{eq19}) into (\ref{eq18}), we can obtain
\begin{equation}
\label{eq20}
\chi T_c x^{d - 1}f^{(d - 1) / (d - 2)}(x) = Af(x)x^{d - 2}(1 + x) -
B\frac{1 - x^d}{1 - x},
\end{equation}
where $x = v_1 / v_2 $, $T_0 = \chi T_c $, $T_c $ is critical temperature.
The value of $\chi=\frac{T}{T_c}$ is form $0$ to $1$. When $x\to 1$ and $\chi \to 1$ ,
the corresponding state is critical state
\begin{equation}
\label{eq21}
f(1) = \frac{d(d - 1)B}{2A}.
\end{equation}
So, the critical point satisfies
\begin{equation}
\label{eq22}
v_2^{d - 2} = v_1^{d - 2} = v_c^{d - 2} = \frac{d(d - 1)B}{2A} = \frac{ks(n
- 1)^2(2ns - 4s + 1)q^{2s}}{16(n - 2)(2s - 1)^2}.
\end{equation}
Substituting eq.(\ref{eq22}) into the eqs.(\ref{eq18}) and (\ref{eq17}), we can obtain
\ba
T_c &=& \frac{2A(d - 2)}{(d - 1)}\left( {\frac{2A}{d(d - 1)B}} \right)^{1 / (d
- 2)} \no  \\ &=& \frac{4(n - 2)(ns - 3s + 1)}{\pi (n - 1)(2ns - 4s + 1)}\left(
{\frac{ks(n - 1)^2(2ns - 4s + 1)q^{2s}}{16(n - 2)(2s - 1)^2}}
\right)^{\textstyle{{1 - 2s} \over {2(ns - 3s + 1)}}} \no ,
\ea
\ba
\label{eq23}
P_c &=& \frac{A(d - 2)}{d}\left( {\frac{2A}{d(d - 1)B}} \right)^{2 / (d - 2)} \no \\
&=& \frac{(n - 2)(ns - 3s + 1)}{\pi s(n - 1)^2}\left( {\frac{ks(n - 1)^2(2ns -
4s + 1)q^{2s}}{16(n - 2)(2s - 1)^2}} \right)^{\textstyle{{1 - 2s} \over {(ns
- 3s + 1)}}}.
\ea
Substituting the eq.(\ref{eq23}) into eq.(\ref{eq20}), we can obtain
\begin{equation}
\label{eq24}
\chi x^{d - 1}f^{(d - 1) / (d - 2)}(x)\frac{2A(d - 2)}{(d - 1)}\left(
{\frac{2A}{d(d - 1)B}} \right)^{1 / (d - 2)} = Af(x)x^{d - 2}(1 + x) -
B\frac{1 - x^d}{1 - x}.
\end{equation}
Because we take account of the case that the temperature $T$ below the critical temperature $T_c$,
, the value of $\chi=\frac{T}{T_c}$ is form $0$ to $1$. When $x\to 1$ and $\chi \to 1$ , the corresponding state
is critical state. For a fixed $\chi $, i.e. a fixed $T_0 $, we can get a certain $x$ from Eq. (\ref{eq24}),
and then according to Eqs. (\ref{eq19}) and (\ref{eq17}), the $v_2 $ and $P_0
$ are solved.
\begin{figure}
  \centering
  \includegraphics[width=4in]{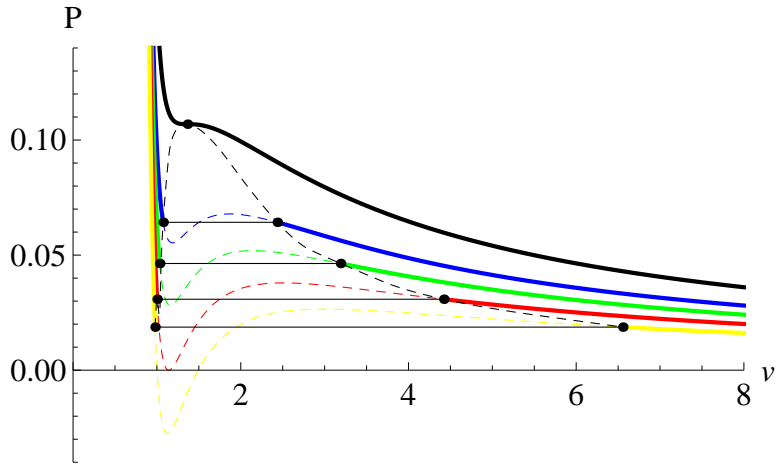}\\
  \caption{\it The simulated isothermal phase transition by isobars and the boundary
of two phase coexistence region for the charged black hole with PMI as
$n=5$, $s=3/4$, $q=0.5$.}\label{coexist}
\end{figure}

To analyze the effect of parameters $n$ and $q$ on the phase
transition processes, we take $\chi =0.1,\;0.3,\;0.5,\;0.7,\;0.9$, and
calculate the quantities $x$, $v_2 $, $P_0 $ as $n =3,\;5,\;6$ and
$q=0.2,\;0.5,\;1$ when $s=3/4$, $s=2$, respectively. The results are shown in
Table \ref{tab1} and \ref{tab2}.
\begin{table}[htbp]
\caption{\it For $s=3/4$, State quantities at phase transition endpoints with different parameters $q$ and spacetime dimensional $n$}
\begin{center}
\begin{tabular}{|p{20pt}<{\centering}|p{20pt}<{\centering}|p{40pt}<{\centering}|p{40pt}<{\centering}|p{40pt}<{\centering}|p{40pt}<{\centering}|p{40pt}<{\centering}|p{40pt}<{\centering}|p{40pt}<{\centering}|p{40pt}<{\centering}|p{40pt}<{\centering}|}
\hline
& &
\multicolumn{3}{|p{120pt}<{\centering}|}{$n =3$} &
\multicolumn{3}{|p{120pt}<{\centering}|}{$n =5$} &
\multicolumn{3}{|p{120pt}<{\centering}|}{$n =6$}  \\
\hline
$q$&
$\chi $&
$x$&
$v_2 $&
$P_0 $&
$x$&
$v_2 $&
$P_0 $&
$x$&
$v_2 $&
$P_0 $ \\
\hline
\raisebox{-6.00ex}[0cm][0cm]{$0.2$}&
0.9&
0.4755&
4.4110&
0.1120&
0.5954&
1.6638&
0.1126&
0.6267&
1.2719&
0.2009\\
\cline{2-11}
 &
0.7&
0.2182&
7.7148&
0.0062&
0.3240&
2.7803&
0.0610&
0.3533&
2.0815&
0.1115 \\
\cline{2-11}
 &
0.5&
0.0859&
17.8745&
0.0022&
0.1491&
5.7145&
0.0246&
0.1678&
4.1802&
0.0459  \\
\cline{2-11}
 &
0.3&
0.0166&
86.3399&
0.0003&
0.0378&
21.65&
0.0046&
0.0449&
15.0825&
0.0090 \\
\cline{2-11}
 &
0.1&
0.00002&
68530.6&
1.41E-7&
0.0002&
4269.4&
8.54E-6&
0.0003&
2267.96&
0.00002 \\
\hline
\raisebox{-6.00ex}[0cm][0cm]{$0.5$}&
0.9&
0.4755&
5.8389&
0.0060&
0.5954&
1.9090&
0.0845&
0.6267&
1.4137&
0.1626  \\
\cline{2-11}
 &
0.7&
0.2182&
10.878&
0.0031&
0.3240&
3.1900&
0.04635&
0.3533&
2.3135&
0.0902 \\
\cline{2-11}
 &
0.5&
0.0859&
25.2035&
0.0011&
0.1491&
6.5565&
0.0187&
0.1678&
4.6464&
0.0372\\
\cline{2-11}
 &
0.3&
0.0166&
121.742&
0.0002&
0.0378&
24.816&
0.0035&
0.0449&
16.7645&
0.0072\\
\cline{2-11}
 &
0.1&
0.00002&
96630&
7.14E-8&
0.0002&
4898.4&
6.48E-6&
0.0003&
2520.88&
0.00002\\
\hline
\raisebox{-6.00ex}[0cm][0cm]{$1$}&
0.9&
0.4755&
7.5721&
0.0035&
0.5954&
2.1181&
0.0687&
0.6267&
1.5314&
0.1386 \\
\cline{2-11}
 &
0.7&
0.2182&
14.107&
0.0018&
0.3240&
3.5395&
0.0377&
0.3533&
2.5062&
0.0769 \\
\cline{2-11}
 &
0.5&
.00859&
32.685&
0.0007&
0.1491&
7.2749&
0.0152&
0.1678&
5.0333&
0.0317\\
\cline{2-11}
 &
0.3&
0.0166&
157.879&
0.0001&
0.0378&
27.5302&
0.0028&
0.0449&
18.16&
0.0062\\
\cline{2-11}
 &
0.1&
0.00002&
125314&
4.25E-8&
0.0002&
5433.1&
5.26E-6&
0.0003&
2730.77&
0.00001 \\
\hline
\end{tabular}
\label{tab1}
\end{center}
\end{table}

\begin{table}[htbp]
\caption{\it For $s=2$, State quantities at phase transition endpoints with different parameters $q$ and spacetime dimensional $n$}
\begin{center}
\begin{tabular}{|p{20pt}<{\centering}|p{20pt}<{\centering}|p{40pt}<{\centering}|p{40pt}<{\centering}|p{40pt}<{\centering}|p{40pt}<{\centering}|p{40pt}<{\centering}|p{40pt}<{\centering}|p{40pt}<{\centering}|p{40pt}<{\centering}|p{40pt}<{\centering}|}
\hline
& &
\multicolumn{3}{|p{120pt}<{\centering}|}{$n =3$} &
\multicolumn{3}{|p{120pt}<{\centering}|}{$n =5$} &
\multicolumn{3}{|p{120pt}<{\centering}|}{$n =6$}  \\
\hline
$q$&
$\chi $&
$x$&
$v_2 $&
$P_0 $&
$x$&
$v_2 $&
$P_0 $&
$x$&
$v_2 $&
$P_0 $ \\
\hline
\raisebox{-6.00ex}[0cm][0cm]{$0.2$}&
0.9&
0.2817&
0.00002&
3.9E8&
0.4513&
0.1033&
28.5&
0.4961&
0.3023&
3.61469\\
\cline{2-11}
 &
0.7&
0.0815&
0.00005&
1.6E8&
0.1985&
0.1980&
14.4142&
0.2355&
0.5508&
1.8853 \\
\cline{2-11}
 &
0.5&
0.0200&
0.0002&
3.9E7&
0.0750&
0.4737&
5.0536&
0.0957&
1.2447&
0.6968   \\
\cline{2-11}
 &
0.3&
0.0016&
0.0018&
2.4E6&
0.0135&
2.4421&
0.6720&
0.0195&
5.7158&
0.1047 \\
\cline{2-11}
 &
0.1&
2.4E-8&
114.59&
12.9982&
0.00001&
2798.52&
0.0002&
0.00003&
3405.2&
0.00006 \\
\hline
\raisebox{-6.00ex}[0cm][0cm]{$0.5$}&
0.9&
0.2817&
0.0045&
6472.47&
0.4513&
0.3100&
3.1617&
0.4961&
0.6632&
0.7514  \\
\cline{2-11}
 &
0.7&
0.0815&
0.0117&
2658.28&
0.1985&
0.5945&
1.5986&
0.2355&
1.2080&
0.3919\\
\cline{2-11}
 &
0.5&
0.0200&
0.0404&
663461&
0.0750&
1.4226&
0.5605&
0.0957&
2.7298&
0.1448 \\
\cline{2-11}
 &
0.3&
0.0016&
0.4365&
41.1060&
0.0135&
7.3332&
0.0745&
0.0195&
12.5362&
0.0217\\
\cline{2-11}
 &
0.1&
2.4E-8&
27247.8&
0.0002&
0.00001&
8403.42&
0.00002&
0.00003&
7469.43&
0.00001\\
\hline
\raisebox{-6.00ex}[0cm][0cm]{$1$}&
0.9&
0.2817&
0.2864&
1.5802&
0.4513&
0.7122&
0.5990&
0.4961&
1.2013&
0.229 \\
\cline{2-11}
 &
0.7&
0.0815&
0.7466&
0.6489&
0.1985&
1.3658&
0.3029&
0.2355&
2.1883&
0.1194\\
\cline{2-11}
 &
0.5&
0.0200&
2.5863&
0.162&
0.0750&
3.2682&
0.1062&
0.0957&
4.9490&
0.0441\\
\cline{2-11}
 &
0.3&
0.0016&
27.9351&
0.01&
0.0135&
16.8472&
0.0141&
0.0195&
22.7087&
0.0066\\
\cline{2-11}
 &
0.1&
2.4E-8&
1.74E6&
5.47E-8&
0.00001&
19306&
4.35E-6&
0.00003&
13530&
3.98E-6 \\
\hline
\end{tabular}
\label{tab2}
\end{center}
\end{table}
From Table \ref{tab1} and \ref{tab2}, it can be seen that $x$ is unrelated to $q$, but incremental with
the increase of $\chi (n)$ at certain $n (\chi)$.
$v_2$ decrease with the increase $\chi$ and $n$ with $s=3/4$.
However, when $s=2$, $v_2$ decrease with the increase $\chi$ and
is nonmonotonic with $n$. $P_0$ increases with the incremental $\chi (n)$ and decreases with the increasing $q$.
The doubt is whether $P_0$ is negative when the temperature
is low enough.

\section{Maxwell equal area law: Some examples}

\label{Cano}

In order to further study the phase transition for the charged black hole with PMI,
we expand the method of Maxwell's equal-area law to the canonical ensemble and
grand canonical ensemble.

\subsection{BTZ-like black holes}
one can select an ensemble in which black hole charge is fixed at infinity. Considering the
fixed charge as an extensive parameter, the corresponding ensemble
is called a canonical ensemble. Interestingly, for $s=n/2$, the solutions (the so-called BTZ
black holes) have different properties. As we will see, for $s=n/2$ the charge term in metric
function is logarithmic and the electromagnetic field is proportional to $r^{-1}$(logarithmic gauge
potential). In other words, in spite of some differences, this special higher dimensional solution
has some similarity withe the charged BTZ solution and reduces to the original BTZ black hole for
$n=2$~\cite{Heidi}.

\begin{figure}[tbp]
$%
\begin{array}{cc}
\epsfxsize=7cm \epsffile{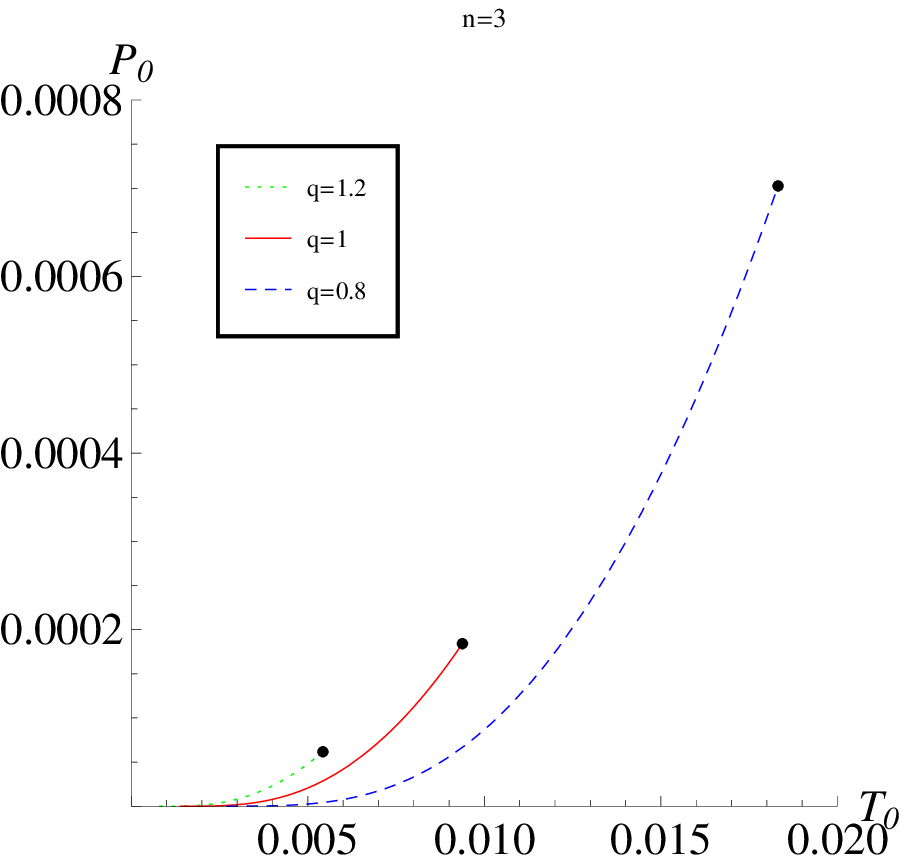} & \epsfxsize=7cm \epsffile{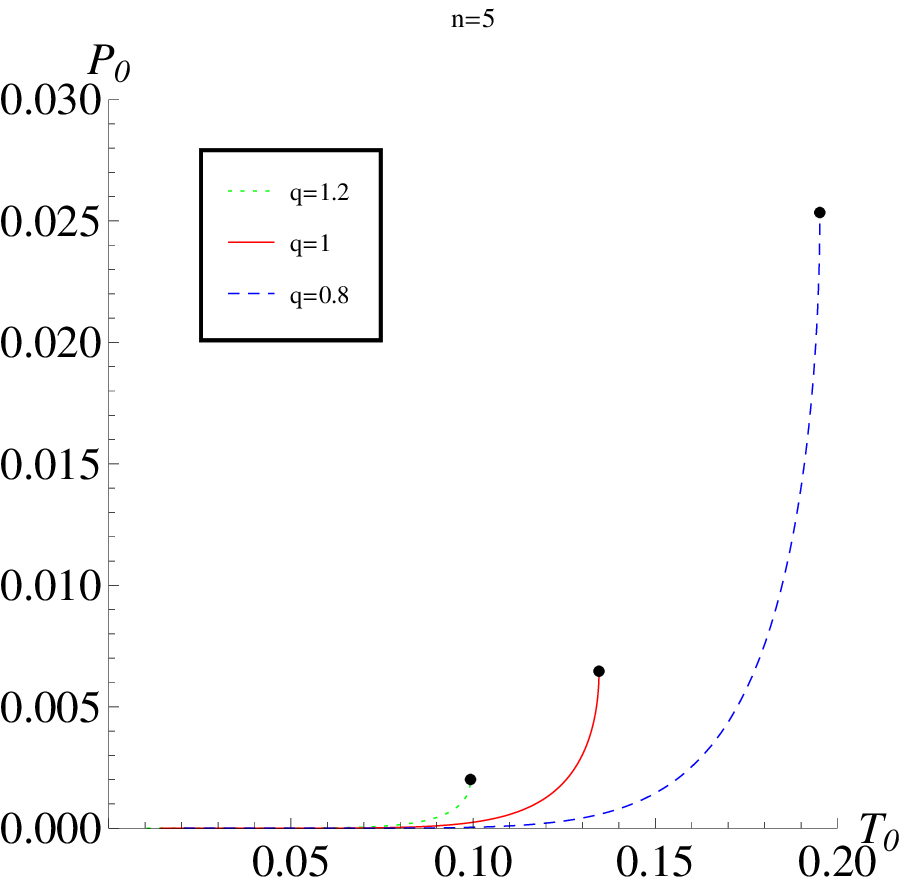}%
\end{array}
$%
\caption{$P-T$ diagram of BTZ-like black holes for $n=3$ (left)
and $n=5$ (right) for the different charge $q$. }
\label{FPVbtz}
\end{figure}

Considering the metric (\ref{Metric}) and the field equations of the bulk action
(\ref{Action}) with $s = n / 2$, we can find that the metric function $f(r)$ and the
gauge potential may be written as~\cite{Heidi}
\begin{equation}
\label{eq25}
f(r) = 1 + \frac{r^2}{l^2} - \frac{m}{r^{n - 2}} - \frac{2^{n / 2}q^n}{r^{n
- 2}}\ln \left( {\frac{r}{l}} \right),
\end{equation}
\begin{equation}
\label{eq26}
A = q\ln \left( {\frac{r}{l}} \right)dt,
\end{equation}
Straightforward calculation show that BTZ-like spacetime has a curvature
singularity located at $r = 0$ in which covered with an event horizon. The
temperature of this black hole be obtained as
\begin{equation}
\label{eq27}
T = \frac{n - 2}{4\pi r_ + }\left( {1 + \frac{n}{n - 2}\frac{r_ + ^2 }{l^2}
- \frac{2^{n / 2}q^n}{(n - 2)r_ + ^{n - 2} }} \right).
\end{equation}
Substituting eq.(\ref{eq7}) into eq.(\ref{eq27}), we can obtain
\begin{equation}
\label{eq28}
P = \frac{T}{v} - \frac{(n - 2)}{\pi (n - 1)v^2} + \frac{1}{16\pi
}\frac{k'q^n}{v^n},
\end{equation}
where
\begin{equation}
\label{eq29}
k' = \frac{2^{5n / 2}}{(n - 1)^{n - 1}},
\quad
v = \frac{4}{(n - 1)}r_ + .
\end{equation}
Substituting eq.(\ref{eq28}) into eq.(\ref{eq13}), we can obtain
\begin{equation}
\label{eq30}
T_0 v_2^{n - 1} x^{n - 1} = Av_2^{n - 2} x^{n - 2}(1 + x) - B'\frac{1 -
x^n}{1 - x},
\end{equation}
and
\begin{equation}
\label{eq31}
v_2^{n - 2} = \frac{B'}{A}\frac{n(1 - x^{n - 1})(1 - x) + (n - 1)(1 -
x^n)\ln x}{x^{n - 2}(n - 1)(1 - x)\left( {2(1 - x) + (1 + x)\ln x} \right)}
= f_1 (x),
\end{equation}
with the method which used the above section, with $B' = \frac{k'q^{2s}}{16\pi }$.
Substituting eq.(\ref{eq31}) into (\ref{eq30})
\begin{equation}
\label{eq32}
\chi T_c x^{n - 1}f_1^{(n - 1) / (n - 2)} (x) = Af_1 (x)x^{n - 2}(1 + x) -
B'\frac{1 - x^n}{1 - x},
\end{equation}
when $x \to 1$, from the eq.(\ref{eq31}), we can get
\begin{equation}
\label{eq33}
f_1 (1) = \frac{n(n - 1)B'}{2A},
\end{equation}
So, the critical point meet with
\begin{equation}
\label{eq34}
v_2^{n - 2} = v_1^{n - 2} = v_c^{n - 2} = \frac{n(n - 1)B'}{2A} =
\frac{k'n(n - 1)^2q^n}{32(n - 2)}.
\end{equation}
Combining (\ref{eq34}), (\ref{eq30}) and (\ref{eq28}), we can obtain
\ba
T_c = \frac{2A(n - 2)}{(n - 1)}\left( {\frac{2A}{n(n - 1)B'}} \right)^{1 /
(n - 2)} = \frac{2(n - 2)^2}{\pi (n - 1)^2}\left( {\frac{32(n - 2)}{k'n(n -
1)^2q^n}} \right)^{1 / (n - 2)}, \no
\ea
\begin{equation}
\label{eq35}
P_c = \frac{A(n - 2)}{n}\left( {\frac{2A}{n(n - 1)B'}} \right)^{2 / (n - 2)}
= \frac{(n - 2)^2}{\pi n(n - 1)}\left( {\frac{32(n - 2)}{k'n(n - 1)^2q^n}}
\right)^{2 / (n - 2)}.
\end{equation}
Combining (\ref{eq35}) and (\ref{eq32}), we can get
\begin{equation}
\label{eq36}
\chi x^{n - 1}f_1^{(n - 1) / (d - 2)} (x)\frac{2A(n - 2)}{(n - 1)}\left(
{\frac{2A}{n(n - 1)B'}} \right)^{1 / (n - 2)} = Af_1 (x)x^{n - 2}(1 + x) -
B'\frac{1 - x^n}{1 - x}.
\end{equation}
We plot the $P-T$ curves with $0<x\le 1$ in Fig.\ref{FPVbtz} when the parameters
$n$, $s$, $q$ take different values respectively. The curves
represent two-phase equilibrium condition for the
charged AdS black hole and the terminal points of the
curves represent corresponding critical points. From fig.\ref{FPVbtz} it can
be seen that the influence of the electric charge $q$ and spacetime $n$ on the
phase diagrams, however, pressure $P_0$ tends zero with decreasing temperature
$T_0$ for all of the fixed $q$ and $n$ cases. The process of phase transition becomes
longer as the spacetime dimensional $n$ is increase. That the pressure $P_0$ is always
positive means Maxwell's equal area law is appropriate to resolve the doubts about
the negative pressure and unstable states in the phase transition of the BTZ-like
black hole.

\subsection{Grand canonical ensemble}
In addition to canonical ensemble, one can work with a fixed electric
potential at infinity. The ensemble of this fixed intensive quantity
translates into the grand canonical ensemble. It is worthwhile to note that,
for linear Maxwell field, the criticality cannot happen in the grand
canonical ensemble~\cite{David,Dolan2}

In this section, we study the critical behavior of charged black holes in
the grand canonical (fixed $\Phi )$ ensemble. We take $q = \Phi r_ + ^{(n -
2s) / (2s - 1)} $ with $v = \frac{4r_ + }{n - 1}$ to rewrite Eq. (\ref{eq10}) in
the following form
\begin{equation}
\label{eq37}
P = \frac{T}{v} - \frac{(n - 2)}{\pi (n - 1)v^2} + \frac{2s - 1}{16\pi }\left(
{\frac{4\sqrt 2 (n - 2s)\Phi }{(2s - 1)(n - 1)v}} \right)^{2s},
\end{equation}
Using the method in the above section, substituting Eq.(\ref{eq37}) into Eq. (\ref{eq13}), we can
obtain
\begin{equation}
\label{eq38}
T_0 v_2^{2s - 1} x^{2s - 1} = Av_2^{2s - 2} x^{2s - 2}(1 + x) - B''\frac{1 -
x^{2s}}{1 - x},
\end{equation}
and
\begin{equation}
\label{eq39}
v_2^{2s - 2} = \frac{B''}{A}\frac{2s(1 - x^{2s - 1})(1 - x) + (2s - 1)(1 -
x^{2s})\ln x}{x^{2s - 2}(2s - 1)(1 - x)\left( {2(1 - x) + (1 + x)\ln x}
\right)} = f_2 (x),
\end{equation}
where $B'' = \frac{2s - 1}{\pi }\left( {\frac{4\sqrt 2 (n - 2s)\Phi }{(2s -
1)(n - 1)}} \right)^{2s}$. Substituting (\ref{eq39}) into (\ref{eq38}), we have
\begin{equation}
\label{eq40}
\chi T_c x^{2s - 1}f_2^{(2s - 1) / (2s - 2)} (x) = Af_1 (x)x^{2s - 2}(1 + x)
- B''\frac{1 - x^{2s}}{1 - x},
\end{equation}
When $x \to 1$, from (\ref{eq39}), we can obtain
\begin{equation}
\label{eq41}
f_2 (1) = \frac{2s(2s - 1)B''}{2A},
\end{equation}
So, the critical point satisfy
\begin{equation}
\label{eq42}
v_2^{2s - 2} = v_1^{2s - 2} = v_c^{2s - 2} = \frac{s(2s - 1)B''}{A} =
\frac{4\sqrt 2 (n - 2s)}{(2s - 1)(n - 1)}\left( {\frac{32s(2s - n)^2}{(n -
2)(n - 1)}} \right)^{1 / (2s - 2)}\Phi ^{s / (s - 1)}.
\end{equation}
Applying Eqs. (\ref{eq38}) and (\ref{eq37}) to the states equation, it is
easy to calculate the critical temperature and critical pressure
\[
T_c = \frac{4A(s - 1)}{(2s - 1)}\left( {\frac{A}{s(2s - 1)B''}} \right)^{1 /
(2s - 2)} = \frac{(s - 1)(n - 2)}{\sqrt 2 \pi (n - 2s)}\left( {\frac{(n -
2)(n - 1)}{32s(n - 2s)^2}} \right)^{1 / (2s - 2)}\Phi ^{ - s / (s - 1)},
\]
\begin{equation}
\label{eq43}
P_c = \frac{A(s - 1)}{s}\left( {\frac{A}{s(2s - 1)B''}} \right)^{1 / (s -
1)} = \frac{(s - 1)(2s - 1)^2}{s\pi }\left( {\frac{(n - 1)(n - 2)}{32(n -
2s)^2s^{1 / s}}} \right)^{s / (s - 1)}\Phi ^{ - 2s / (s - 1)}.
\end{equation}
Combining (\ref{eq43}) and (\ref{eq40}) and taking $\chi$ is constant, we find that $x$ satisfy the equation
\begin{equation}
\label{eq44}
\chi x^{2s - 1}f_2^{(2s - 1) / (2s - 2)} (x)\frac{2A(2s - 2)}{(2s -
1)}\left( {\frac{A}{s(2s - 1)B''}} \right)^{1 / (2s - 2)} = Af_2 (x)x^{2s -
2}(1 + x) - B''\frac{1 - x^{2s}}{1 - x}.
\end{equation}

For a fixed $\chi $, i.e. a fixed $T_0 $, we can get a certain
$x$ from Eq. (\ref{eq44}), and then according to Eqs. (\ref{eq37}) and (\ref{eq39}), the $v_2 $ and $P_0
$ are solved. The corresponding $v_1 $ can be got from $x=v_1 /v_2 $.
Join the points $(v_1 ,P_0 )$ and $(v_2 ,P_0 )$ on isotherms in $P-v$
diagram, which generate an isobar representing the process of isothermal
phase transition or the two phase coexistence situation like that of van der
Waals system. Fig.4 shows the isobars on the background of isotherms at
different temperature and the boundary of the two-phase equilibrium
region by the dot-dashed curve as $n=3$, $s=6/5$ and $n=4$, $s=5/4$, respectively.
The isothermal phase transition process becomes shorter as the temperature
goes up until it turns into a single point at a certain temperature, which
is critical temperature, and the point corresponds to critical state of the
charged AdS black hole with PMI in the grand canonical ensemble.
\begin{figure}[tbp]
$%
\begin{array}{cc}
\epsfxsize=7cm \epsffile{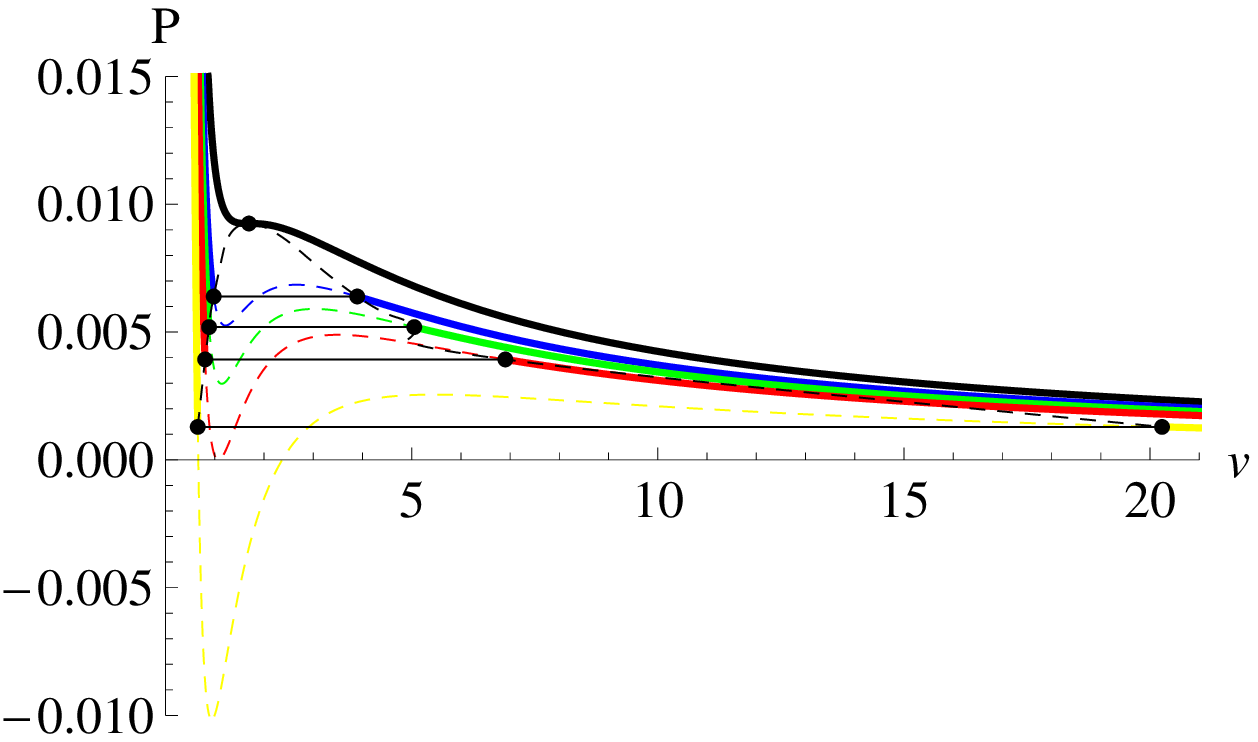} & \epsfxsize=7cm %
\epsffile{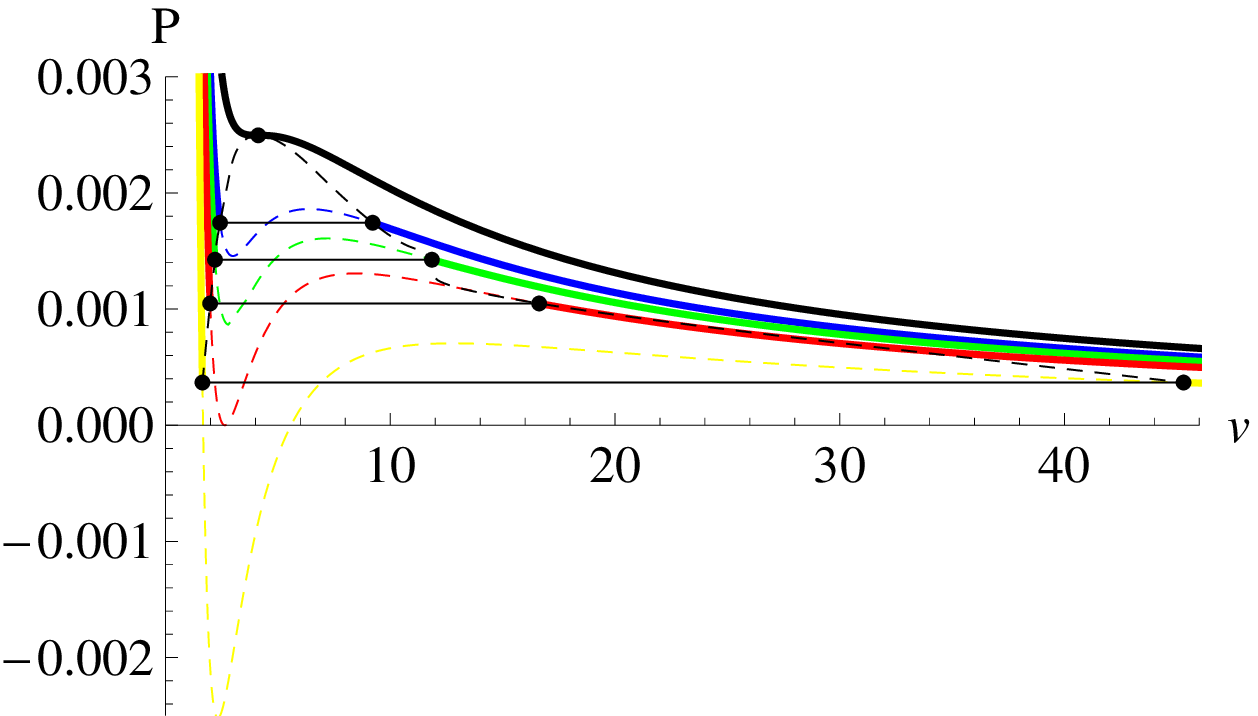}%
\end{array}
$%
\caption{$P-v$ diagram of charged AdS black holes in PMI for
$s=\frac{6}{5}$ with $n=3$ (left) and $s=\frac{5}{4}$ with $n=4$
(right). The temperature of isotherms decreases from top to
bottom. The bold line is the critical isotherm diagram.}
\label{PotPV}
\end{figure}

\section{Concluding Remarks}
In this paper we have extended the idea of fluid/gravity analog
in order to provide a new picture of the isothermal behavior
of critical charged black hole in AdS background with a nonlinear source.
The results of this method is that physical black hole undergoes
an isothermal transition from gas to liquid phase at constant
pressure. Consequently there are neither regions with negative
nor divergent specific heat. Furthermore, we were able to obtain
analytic solutions the area law both in canonical(grand canonical) ensemble.
We conclude that working in the $(P,v)$ plane gives non-trivial advantage
with respect to the Van der Walls description in $(P,V)$ plane.

The charged AdS black hole with PMI is regarded as a
thermodynamic system, and its state equation has been derived.
But when temperature is below critical temperature,
thermodynamic unstable situation appears on isotherms, and when
temperature reduces to a certain value the negative pressure
emerges, which can be seen in Fig.1, Fig.2 and Fig.4. However, by Maxwell
equal law we established an phase transition process and the problems can be
resolved. The phase transition process at a defined temperature
happens at a constant pressure, where the system specific
volume changes along with the ratio of the two coexistent phases.
According to Ehrenfest scheme the phase transition belongs to the
first order one , which can be seen in Table 1 and Table 2. We draw the isothermal
phase transition process and depict the boundary of two-phase
coexistence region in Fig.4. The obtained $P_0-T_0$(Fig.3) diagram for
different dimensions $n$ shows that as dimensionality increases, the temperature of
critical points increases which indicates the necessity of more energy for having a phase transition.

Taking black hole as an thermodynamic systems, many investigations show
the phase transition of some black holes in AdS spacetime and dS
spacetime is similar to that of van der Waals-Maxwell gas-liquid
system~\cite{Cai}, and the phase transition of some
other AdS black hole is alike to that of multicomponent superfluid
or superconducting system\cite{Altami,Altami1,Altami2}. It would make sense
if we can seek some observable system, such as van der Waals
gas, to back analyze physical nature of black holes by their similar
thermodynamic properties. That would help to further understand the
thermodynamic quantities, such as entropy, temperature, heat capacity and
so on, of black hole and that is significant for improving
self-consistent thermodynamic theory of black holes. Also, we have applied
the same procedure for the BTZ-like black
holes to obtain their phase transition. Calculations showed that
thermodynamic behaviors of BTZ-like black holes are the same as
PMI ones. Moreover, we have studied the grand canonical ensemble in which
the potential, instead of charge, should be fixed on the boundary.
In contrast to the Maxwell case, here one sees a
phase transition. Finally, Perhaps a holographic approach helps us to have a
better understanding of this problem. We leave the study of these
interesting questions for future studies.

\section{Acknowledgement}

The authors are grateful to Meng-Sen Ma for his valuable discussions.
 This work was supported by the Young Scientists Fund of the National
Natural Science Foundation of China (Grant No.11205097), in part by
the National Natural Science Foundation of China (Grant Nos.11475108),
Supported by Program for the Innovative Talents of Higher Learning Institutions of Shanxi,
the Natural Science Foundation of Shanxi Province,China(Grant No.201601D102004)
and the Natural Science Foundation for Young Scientists of Shanxi Province,China
(Grant No.2012021003-4),the Natural Science Foundation of Datong city(Grant No.20150110).

%%%%%%%%%%%%%%%%%%%%%%%%%%%%%%%%%%%%%%%%%%%%%%%%%%%%%%%%%%%%%%%%%%%%%%%%%%%%%%%%%%%%%%%%%%%%%%%%%%%%%%%%%%%%%%%%%%%%%%%%%%%%%%%%%%%%%%%%%
%%%%%%%%%%%%%%%%%%%%%%%%%%%%%%%%%%%%%%%%%%%%%%%%%%%%%%%%%%%%%%%%%%%%%%%%%%%%%%%%%%%%%%%%%%%%%%%%%%%%%%%%%%%%%%%%%%%%%%%%%%%%%%%%%%%%%%%%%%

\end{document}